\documentclass[conference]{IEEEtran}
\IEEEoverridecommandlockouts
\usepackage{cite}
\usepackage{amsmath,amssymb,amsfonts}
\usepackage{multirow}
\usepackage{algorithmic}
\usepackage{graphicx}
\usepackage{textcomp}
\usepackage{xcolor}
\usepackage{amsmath}
\usepackage{url}
\usepackage{balance}
\makeatletter
\newcommand{\verbatimfont}[1]{\renewcommand{\verbatim@font}{\ttfamily#1}}
\makeatother
\verbatimfont{\tiny}

\def\BibTeX{{\rm B\kern-.05em{\sc i\kern-.025em b}\kern-.08em
    T\kern-.1667em\lower.7ex\hbox{E}\kern-.125emX}}

\begin{document}

\title{QUANTIFY: A framework for resource analysis and design verification of quantum circuits}

\author{\IEEEauthorblockN{Oumarou Oumarou}
\IEEEauthorblockA{\textit{Department of Informatics}\\
\textit{Clausthal University of Technology},  \\
38678 Clausthal-Zellerfeld, Germany}
\and
\IEEEauthorblockN{Alexandru Paler}
\IEEEauthorblockA{\textit{Institute for Integrated Circuits}\\
\textit{Johannes Kepler University}, \\
4040 Linz, Austria}
\and
\IEEEauthorblockN{Robert Basmadjian}
\IEEEauthorblockA{\textit{Department of Informatics}\\
\textit{Clausthal University of Technology}, \\
38678 Clausthal-Zellerfeld, Germany}
}

\maketitle

% \makeatletter
% \def\ps@IEEEtitlepagestyle{
%   \def\@oddfoot{\mycopyrightnotice}
%   \def\@evenfoot{}
% }

\setlength{\abovecaptionskip}{-0.0mm}

\begin{abstract}
Quantum resource analysis is crucial for designing quantum circuits as well as assessing the viability of arbitrary (error-corrected) quantum computations. To this end, we introduce QUANTIFY, which is an open-source framework for the quantitative analysis of quantum circuits. It is based on Google Cirq and is developed with Clifford+T circuits in mind, and it includes the necessary methods to handle Toffoli+H and more generalised controlled quantum gates, too. Key features of QUANTIFY include: (1) analysis and optimisation methods which are compatible with the surface code, (2) choice between different automated (mixed polarity) Toffoli gate decompositions, (3) semi-automatic quantum circuit rewriting and quantum gate insertion methods that take into account known gate commutation rules, and (4) novel optimiser types that can be combined with different verification methods (e.g. truth table or circuit invariants like number of wires).  For benchmarking purposes QUANTIFY includes quantum memory and quantum arithmetic circuits. Experimental results show that the framework's performance scales to circuits with thousands of qubits.
%maybe tens of thousands?
\end{abstract}

\maketitle

% Call for papers
% http://www.eng.ucy.ac.cy/theocharides/isvlsi20/ISVLSI-2020_QCW.pdf

\section{Introduction}
\label{sec:intro}

Quantum computing promises to solve practical problems from chemistry, biology, cryptography in a more efficient and swiftly manner than traditional computing. Having demonstrated that quantum computational supremacy is achievable \cite{arute2019quantum}, the next step is to perform the first error-corrected experiments in order to pave the way to large scale error-corrected quantum computations using for instance NISQ (Noisy Intermediate-Scale Quantum) devices. To this end, a vast amount of research effort is invested in the technological realisation of quantum computers. Significant progress has been achieved in designing and realising quantum computers \cite{arute2019quantum} as well as implementing quantum algorithms \cite{coles2018quantum}.

Analogous to traditional computing paradigm, the quantum algorithms can be expressed as quantum circuits. And similarly to the classical world, efficient circuit design is one of the most challenging phases during the realisation of a computing system. Additional costs are incurred if a circuit's design is not the result of a thorough and correct analysis of the algorithm that is going to be executed. In the case of quantum computing, quantum circuit design methods have received considerable attention. Nevertheless, realistic quantum resource estimations \cite{paler2019really} and circuit optimisations are of vital importance for industrially relevant quantum computations. It is within this context that, to the best of our knowledge, there is no generic, robust and scalable software framework to support synthesis, analysis, optimisation and verification of quantum circuits with respect to cost metrics relevant to non-error-corrected and error-corrected computations.

\begin{figure}[t!]
    % the source of this image quantify.drawio can be edited with https://app.diagrams.net/
    \centering
    \includegraphics[scale=0.25]{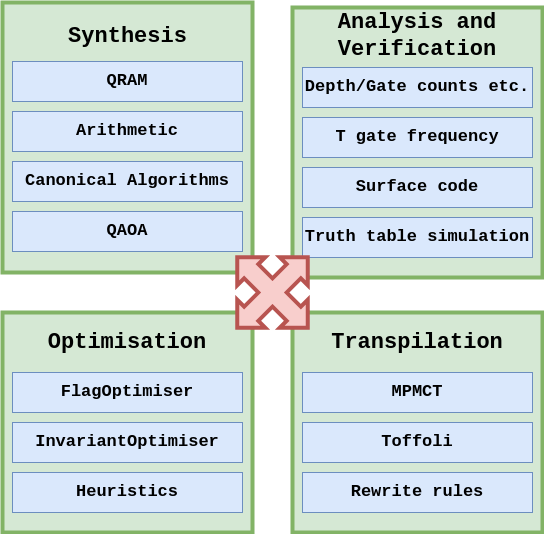}
    \caption{The four step types of QUANTIFY can be executed almost in any order by forming a pipeline workflow.}
    \label{fig:workflow}
\end{figure}

This paper introduces QUANTIFY, which is an open-source platform with an agile and robust design. QUANTIFY paves the way towards a more efficient and scalable automated design tool for quantum circuits, because its most important features are: (1) the capability to perform quantum circuit optimisations driven by the insights obtained through circuit analysis, (2) optimisation methods that are coupled to feedback loops that include verification procedures, (3) flexible gate transpilation mechanisms, in particular for reversible gate decomposition, and (4) analysis and optimisation methods designed specifically for quantum circuits protected by the quantum surface error-correcting codes. Moreover, QUANTIFY can be applied to existing circuits, and it can be used to synthesise new circuits, as well.

The rest of this paper is organised as follows: In Section~\ref{sec:quantify}, the steps of circuit synthesis, transpilation, optimisation, and verification as well as analysis are presented. The implementation details of QUANTIFY by considering the step-wise design are given in Section~\ref{sec:implementation}. The experimental results are demonstrated in Section~\ref{sec:results}. Section~\ref{sec:conclusion} concludes the paper.

\section{QUANTIFY}\label{sec:quantify}

QUANTIFY is an open-source framework\footnote{\url{https://github.com/quantumresource/quantify}} for the quantitative (e.g. numerical) resource analysis of quantum circuits. From a practical and realistic perspective, the resources required to execute a quantum circuit, with or without error-correction, include the number of physical qubits. The number of physical qubits is determined by the so-called width of the computation \cite{paler2019really}. The amount of time for operating the physical qubits is determined by the depth of the computation. The depth time is also a reflection of execution performance. Moreover, both the number of physical qubits (width) and execution performance (depth) influence on the total energy consumed to perform the computation \cite{arute2019quantum}.

The resource analysis results are relevant only if the circuits under study are correct. Consequently, analysis has to be supported by methods that guarantee a certain degree of correctness for the quantum circuit design and implementation. It has to be ensured that the analysed circuits are conforming to a given design specification. For example, when optimising a quantum adder, the circuit obtained after each optimisation step should be correct with respect to some criteria. As a result, QUANTIFY was developed to include novel methods to achieve a flexible array of verification criteria which range from functional (e.g. truth table) to structural (e.g. circuit-level characteristics) conditions.

QUANTIFY fulfils two requirements: from a design perspective, it is agile and generic, and from an execution perspective it is responsive and robust. Agility and generality are reflected in the fact that new features and methods, as well as quantum circuits for benchmarking purposes, can be added on the fly. Responsiveness and robustness are supported within the framework by carrying out the analysis of the circuits in a reasonable amount of time (even for very large numbers of qubits, as illustrated in the Results section), and the correctness of the outputs can be validated in the sense of cross-checked against a specification.

To this end, the architecture of QUANTIFY is designed in a, by now classical, workflow manner that consists of four step types (e.g. Fig.~\ref{fig:workflow}): (1) circuit synthesis, (2) gate level transpilation (e.g. translation from one gate set into another), (3) circuit optimisation and (4) analysis and verification. The advantage of such a step-wise workflow-design is the fact that each step (a) is independent of the others, (b) can be enhanced with new features, and (c) its correctness and execution time can be adequately validated and computed respectively. The steps are designed such that these can be arranged in a  processing pipeline: the output of one step serves as input to another step. Thus, during a quantum circuit analysis and optimisation procedure, each of the four step types can be repeated multiple times and can be executed in an arbitrary order. Each step type is described in the following.

\subsection{Step 1: Circuit Synthesis}

QUANTIFY can be applied to any Cirq circuit, meaning that the entire circuits and examples library of Cirq can be used for analysis and verification. Moreover, QUANTIFY includes state-of-the-art synthesis methods for quantum random access memory (QRAM) circuits, such as Bucket Brigade (BB), Large Depth Small Width (LDSW), and Small Depth Large Width (SDLW). Arithmetic circuits, such as adders and multipliers, are included with QUANTIFY as well.

\subsection{Step 2: Gate Level Transpilation}

Once a circuit is synthesised, its gates can be transpiled from one quantum gate set to another. For example, arithmetic circuits are very often formulated using Toffoli+H gates or even more general MPMCT (Mixed Polarity Multiple Control Toffoli) gates. Transpilation is in general performed from one gate set into Clifford+T, because on the one hand that is the gate set that captures most of the physical quantum hardware instructions, and on the other hand those gates are supported by the quantum error-correcting codes (e.g. the surface code). Consequently, QUANTIFY provides various methods of transpilation into Clifford+T gates. In the following, the terms decomposition and transpilation are used interchangeably.

Toffoli gates are transpiled into Clifford+T by using phase polynomial decompositions of the Toffoli gate \cite{selinger2013quantum}. QUANTIFY includes, to the best of our knowledge, the most relevant state-of-the-art and frequently used Toffoli gate decompositions, such as the ones from \cite{selinger2013quantum} and \cite{amy2014polynomial}. MPMCT transpilation is based on the classic method from \cite{barenco1995elementary}: every MPMCT of $n$ qubits and $m$ control qubits can be  equivalently presented by a circuit consisting of $4(m-2)$ Toffoli gates. Once Toffoli gates are obtained, their Clifford+T decomposition is chosen from the dictionary of supported ones.

\subsection{Step 3: Circuit Optimisation}

The output from previous step, which is the decomposed gate level presentation of the original quantum circuit of Step 1, is used either to apply further optimisations or to perform verification (see Section~\ref{sec:verification}). Thus, in this step reordering as well as removal of unnecessary gates are performed for the sake of optimally presenting the quantum circuit under study. QUANTIFY implements three different types of optimisations mechanism, which are described in Section~\ref{sec:optimisation}.

\subsection{Step 4: Analysis and Verification}
\label{sec:verification}

After generation and gate-level decomposition of a quantum circuit, it is possible either to verify the output from Step 2 or Step 3. For the latter case, this "Verification" step can serve as a feedback loop back to the "Optimisation" step to verify whether the applied optimisations contributed to further improvements. QUANTIFY provides several metrics which allows the quantum circuit under study to be verified, such as the number of Clifford+T (T-count), Hadamard and CNOT gates as well as the depth of the circuit and the Clifford+T gates (T-depth). As mentioned above, it is to be noted that the Steps 2, 3 and 4 of the workflow are not executed sequentially, such that several iterations can happen based on the feedback of the verification metrics provided by Step 4.     

\section{Implementation and Examples}\label{sec:implementation}

The core of QUANTIFY is Google Cirq\footnote{\url{https://github.com/quantumlib/Cirq}} -- a Python-based platform for creating, editing, and invoking Noisy Intermediate Scale Quantum (NISQ) circuits. We have chosen Cirq due to its versatility, support of both high- and low-level compilation, and optimisation methods. The skeleton of QUANTIFY is implemented in four software components corresponding to the design from Fig.~\ref{fig:workflow}.

QUANTIFY advances the state-of-the-art. Table~\ref{tab:my_label} is a comparison between Quipper\cite{quipper}, Qiskit\cite{qiskit}, Q\#\cite{svore2018q}, ProjectQ\cite{projectq} and RevKit\cite{revkit} and the features of QUANTIFY as explained in the following sections. To the best of our knowledge, QUANTIFY is the only framework by now that supports mixing transpilation methods during a circuit pass.

\begin{table}[t!]
    \footnotesize
    \centering
    \caption{Comparison with other quantum software}
    \label{tab:my_label}
    \setlength{\tabcolsep}{4pt} %% default is 6pt
    \begin{tabular}{l|c|c|c|c|c|c}
     & Qiskit & ProjectQ & Q\# & Revkit & Quipper & QUANTIFY\\
    \hline
     Transpilation & X & X & X & X & X & X\\
     Flag Opt. & & & & & & X\\
     Analysis     &  & X & X &  & & X\\
     Invariant Verif.  &  &  &  &  & X & X
     \end{tabular}
\end{table}

\subsection{Circuits}

Circuit synthesis is interconnected with the other components of QUANTIFY. For example, the circuit classes expose a \emph{metrics} interface, that enables basic resource analysis in a manner similar to ProjectQ \cite{projectq} and Q\#\cite{svore2018q}. This interface can determine, among others, the following costs: circuit depth, T-depth, T-, Hadamard-, CNOT-, and qubit-count.

In practice, circuit synthesis scripts are often edited or extended to obtain circuits conforming to a certain specification. It is very useful to automatically ensure that the synthesised circuits are still conforming to a given specification. QUANTIFY's robustness with respect to circuit synthesis is ensured by cross-checking the circuits against cost metrics such as the T-count for instance. This mechanism increases the guarantees that no programming bugs or other types of errors were introduced into the circuits.

The herein presented circuits are not available in Cirq. For benchmarking purposes, QUANTIFY contains the implementation of state-of-the-art QRAM circuits such as Bucket Brigade, Large Depth Small Width (LDSW), and Small Depth Large Width (SDLW). Arithmetic circuits are available, as well. The latter are mostly formulated using reversible gates such as the Toffoli gate, but in practice resource analysis is concerned with the Clifford+T gate set. 

As a result, synthesis and transpilation (see Section~\ref{sec:transpil}) function as a hybrid: QUANTIFY sythesises an instance of the corresponding quantum circuit by calling a generic \textit{constructor}, which takes as a parameter a flag that indicates how to transpile the initial gates of the circuit. The transpilation type (see Section~\ref{sec:transpil}) can also be left unspecified, such that the resulting circuit is not decomposed. Nevertheless, transpilation can be performed at any time, during or after circuit synthesis.

\begin{figure}[h!]
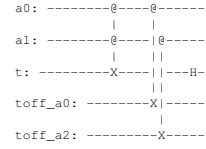

\begin{verbatim}
                          a0: --------@----@------
                                      |    |
                          a1: --------@----|@-----
                                      |    ||
                          t: ---------X----||---H-
                                           ||
                          toff_a0: --------X|-----
                                            |
                          toff_a2: ---------X-----
\end{verbatim}
\caption{A Toffoli gate on the qubits named a0, a1 and t followed by two CNOT gates and a Hadamard gate.}
\label{fig:example01}
\end{figure}

The following sections will use Fig.~\ref{fig:example01} to illustrate the implementation of transpilation, optimisation and verification.

\subsection{Transpilation}
\label{sec:transpil}

Cirq includes a very flexible mechanism to represent quantum gate sets, and this capability is inherited by QUANTIFY. Quantum circuits can be easily transformed from one gate set into another, during a process called transpilation. Some transpilation procedures are easier than others. For example, it is easier to decompose (e.g. express a single gate as a sequence of multiple gates) than to recompose (the inverse operation of a decomposition).

QUANTIFY implements two types of reversible gate decompositions: MPMCT or Toffoli transpilation method. There is flexibility with respect to the choice of the qubit ordering during the decomposition. For example, in the case of Toffoli decompositions, two qubits are controls and their order influences the decompositions and optimisations.

\begin{figure}[h!]
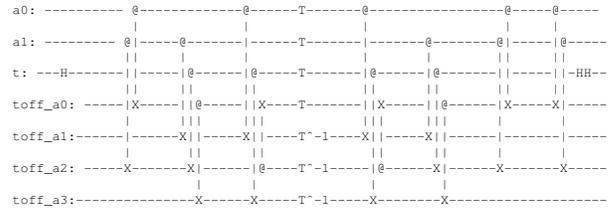

\begin{verbatim}
    a0: ---------- @-------------@------T-------@-----------------@-----@-----
                   |             |              |                 |     |
    a1: --------- @|-----@-------|------T-------|-------@--------@|-----|@-----
                  ||     |       |              |       |        ||     ||
    t: ---H-------||-----|@------|@-----T-------|@------|@-------||-----||-HH--
                  ||     ||      ||             ||      ||       ||     ||
    toff_a0: -----|X-----||@-----||X----T-------||X-----||@------|X-----X|-----
                  |      |||     |||            |||     |||      |       |
    toff_a1:------|------X||-----X||----T^-1----X||-----X||------|-------|-----
                  |       ||      ||             ||      ||      |       |
    toff_a2: -----X-------X|------|@----T^-1-----|@------X|------X-------X-----
                           |      |              |        |      
    toff_a3:---------------X------X-----T^-1-----X--------X--------------------

\end{verbatim}
\caption{The Toffoli was decomposed into Clifford+T gates using the decomposition from \cite{selinger2013quantum}. There are three Hadamard gates on qubit t.}
\label{fig:example02}
\end{figure}

For the circuit example from Fig.~\ref{fig:example01}, if the T-depth of one decomposition of the Toffoli gate is specified and the ordering of controls is a0, a1, then the transpiled circuit is Fig.~\ref{fig:example02}.
 
\subsection{Optimisation: Flags and Commutations}
\label{sec:optimisation}

The optimisation component is responsible for optimising (e.g. minimising) the circuit under study with respect to a variety of cost metrics such as CNOT-count, T-count, gate parallelism, etc. QUANTIFY supports circuit optimisation heuristics by means of gate patterns. 

The framework includes a \emph{FlagOptimiser} class of optimisation strategies which applies circuit identities, also known as templates or rewrite rules, only to gates which were flagged beforehand. The rewrite rules are specified by the user. In QUANTIFY, \emph{CancelCNOT} and \emph{CancelHadamard} are instances of the \emph{FlagOptimiser}. The former cancels adjacent CNOTs and the latter cancels adjacent Hadamard gates (both the CNOT and the Hadamard are their own inverse operation).

The advantage of a \emph{FlagOptimiser} is that it can steer the regions where the optimisations are applied. For example, a preliminary analysis may indicate that a circuit's depth can be minimised by cancelling some neighbouring CNOTs. The CNOT's circuit identity to reduce CNOT-count will be applied automatically only to the CNOTs that were flagged according to a criteria (e.g. applied to a given set of qubits). Such a procedure ensures that the structural changes to the circuit can be tracked through optimisation, and this is a very useful property during design verification.

\begin{figure}[h!]
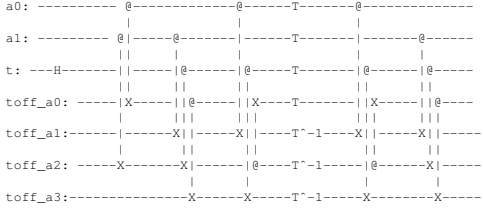

\begin{verbatim}
                a0: ---------- @-------------@------T-------@--------------
                               |             |              |                 
                a1: --------- @|-----@-------|------T-------|-------@------
                              ||     |       |              |       |       
                t: ---H-------||-----|@------|@-----T-------|@------|@-----
                              ||     ||      ||             ||      ||       
                toff_a0: -----|X-----||@-----||X----T-------||X-----||@----
                              |      |||     |||            |||     |||      
                toff_a1:------|------X||-----X||----T^-1----X||-----X||-----
                              |       ||      ||             ||      ||      
                toff_a2: -----X-------X|------|@----T^-1-----|@------X|-----
                                       |      |              |        |      
                toff_a3:---------------X------X-----T^-1-----X--------X-----
\end{verbatim}
\caption{On the right hand side of circuit from Fig.~\ref{fig:example02}, two Hadamards and four CNOT gates were cancelled.}
\label{fig:example03}
\end{figure}

Simultaneously, by attaching flags to gates, it is possible to track how the gates are commuted through the circuit. Tracking supports the diagnosis and debugging of optimisation heuristics. Very recently Cirq included the support for flags on gates, too. Fig.~\ref{fig:example03} illustrates the result of applying \emph{CancelCNOT} and \emph{CancelHadamard} to the circuit from Fig.~\ref{fig:example02}.

\begin{figure}[h!]
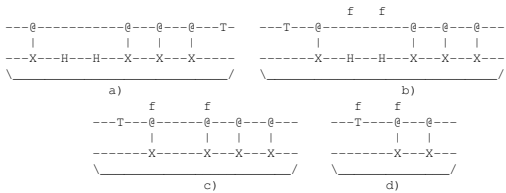

\centering
\begin{verbatim}
                                                    f   f
         ---@-----------@---@---@---T-   ---T---@-----------@---@---@---   
            |           |   |   |               |           |   |   |       
         ---X---H---H---X---X---X-----   -------X---H---H---X---X---X---  
         \___________________________/   \_____________________________/  
                      a)                                   b)                       
                           f      f                  f    f
                    ---T---@------@---@---@---    ---T----@---@---
                           |      |   |   |               |   |
                    -------X------X---X---X---    --------X---X---
                    \________________________/    \______________/
                                  c)                     d)
\end{verbatim}
\caption{Example of using a \emph{CommutationOptimiser} and a \emph{FlagOptimiser}: a) Initial circuit; b) A \emph{CommutationOptimiser} will move the T gate towards the input region; the two Hadamard gates are flagged (cf. the f symbol); c) The \emph{FlagOptimiser} cancels the flagged H gates, and the flag is transferred to the two surrounding CNOT gates. d) The two flagged CNOTs are cancelled and the flag is transferred to the surrounding T and CNOT gates. The two currently flagged gates cannot be cancelled.}
\label{fig:excom01}
\end{figure}

Another type of QUANTIFY optimiser is the \emph{CommutationOptimiser}. It is also based on circuit identities, but the goal is to commute gates, as shown for example in Fig.~\ref{fig:excom01}. A \emph{CommutationOptimiser} tries to commute as many gates of a particular type towards a certain region of the circuit, without cancelling those gates. The \emph{CommuteTGatesToStart} optimiser is based on the circuit commutativity between the control of a CNOT and T gates, and moves the T gates as far as possible to the left hand side of the circuit by commuting them with CNOT controls. Afterwards, a hypothetical \emph{FlagOptimiser} could re-compose neighbouring T gates into S gates.

Combining the \emph{CommutationOptimiser} with a \emph{FlagOptimiser} is useful to determine the correctness of the optimisation heuristics. Such a method helps answering questions such as: were the gates correctly reduced? Why is a particular number of gates being reduced? An example is illustrated in Fig.~\ref{fig:excom01}. 

\subsection{Analysis: Gate Distribution and Resource Estimation}

Circuit analysis is used for both steering optimisation heuristics (such as the \emph{FlagOptimiser}), and also as an integral part of the verification procedures (see following section). For the examples from Fig.~\ref{fig:example02} and Fig.~\ref{fig:example03}, QUANTIFY returns that the number of CNOTs was reduced from 18 to 14.

QUANTIFY includes also more detailed analysis methods, such as the distribution of T gates into the circuit \cite{paler2019clifford}. By analysing how T gates are arranged in a circuit (i.e. the moment and time at which these are applied effectively in the circuit), it is possible to perform optimal scheduling of the T-state distillation procedures imposed by the surface code\cite{paler2019really}.

\subsection{Verification: Exhaustive and Invariant Checking}

The goal of verification is to guarantee that the circuits which were compiled and optimised are correct. The design of a quantum circuit includes many aspects, and correspondingly the verification methods are devised for a wide range of purposes. For the verification of structural properties, QUANTIFY includes the \emph{InvariantCheckingOptimiser}. This kind of optimiser is a unique feature of QUANTIFY, and verifies that a given criteria remains invariant throughout the optimisation procedure. For example, when minimising a circuit's T-count, the counts of other gate types, such as Hadamard, is expected to be an invariant and not change. The InvariantCheckingOptimiser is, compared to the method presented in \cite{hietala2019verified}, a more relaxed type of verified optimisation because it does not use formal proofs to determine the correctness of the optimisation method. Bugs or other problems within the optimisation procedure can be diagnosed if the value of the invariance criteria is changing. For the example of Fig.~\ref{fig:example03}, where the Hadamard- and CNOT-counts are minimised, the invariant is the T-count: at each optimisation step there is a single T gate in the circuit. As mentioned before this functionality is unique to QUANTIFY, compared to the existing quantum frameworks development, which makes the debugging an easier task hence making it an excellent choice especially for scientific purposes.  

Truth table verification is an exhaustive verification method that ensures that the quantum circuit is performing the specified input-output transformations. Due to its exponential complexity, this method does not scale for large circuits, but can be used, for example, for circuit identities within the \emph{Commutation}- and \emph{FlagOptimiser}s. Moreover, truth table verification can be easily applied to reversible circuits, because these are based exclusively on Toffoli gates, which are tightly related to Boolean logic. For the more quantum-computing-like Clifford+T circuits, truth table verification is not straightforward, but methods relying on stabiliser truth tables \cite{paler2018specification}, or probabilistic quantum circuit simulations using tensor network simulators \cite{villalonga2019flexible} may be used.  In QUANTIFY, the exhaustive reversible circuit simulation uses the standard quantum circuit simulator provided by Cirq, and tensor network simulation is provided through the Cirq interface to QFlex \cite{villalonga2019flexible}.

\section{Synthesis and Optimisation of Extremely Large Circuits}
\label{sec:results}

The responsiveness of QUANTIFY was benchmarked for the "Synthesis" and "Optimisation" steps, because these are the most resource intensive ones. We used the QRAM Bucket Brigade circuits from \cite{di2020fault}, due to their interesting property that their number of wires $q$ increases exponentially with linear address lengths $n$, such that $q=n+2^{n+1}+5$. The goal of this section is to empirically show that QUANTIFY has the necessary baseline performance to synthesise and optimise circuits consisting of tens of thousands of qubits.

More performance can be unlocked by improving CPU-responsiveness of the underlying Cirq platform. Interestingly, while monitoring the hardware resources during the execution of the experiments, we observed that QUANTIFY is CPU-intensive (see Sections~\ref{sec:exp_exec_time} and~\ref{sec:exp_mem_usage}). This is in contrast to the assumption that memory would be a bottleneck due to the type of QRAM circuits chosen to investigate. QUANTIFY is a CPU-intensive framework, because it is based on Google Cirq (see Section~\ref{sec:implementation}), which is, for the moment, optimised for ease-of-use instead of speed (e.g. not parallelised). We expect that data structures specialised for quantum circuit representation, such as \cite{paler2019faster},  will improve drastically the overall execution time of QUANTIFY.

\subsection{Configuration}

The experimental evaluation was focused on: a) execution time and b) memory usage. To this end, we use two scenarios: 1) "Synthesis" is performed for QRAM circuits where $2\leq n\leq 19$; 2) "Synthesis \& Optimisation"  is carried out on circuits having $2\leq n\leq 12$.

To the best of our knowledge, QUANTIFY is the first framework shown to support realistic synthesis and optimisation of arbitrary large-scale quantum circuits. Note that for an address size of $n\geq10$ qubits, this corresponds to QRAM circuits of thousands of gates and wires. Previous works, such as \cite{nam2018automated}, included a restricted set of gate level optimisations and the optimisation heuristic was not flexible. As detailed in Section~\ref{sec:optimisation}, QUANTIFY is flexible with respect to optimisation heuristics.

The experiments were executed on an Intel i7-7700K \cite{intel} quad-core machine with 32 GB of RAM, running a Ubuntu 19.04. QUANTIFY has not been parallelised for multi-threading or multi-core technologies. For fair results, we pinned QUANTIFY to one of the cores of the processor (through the Linux command \emph{taskset}), shielded that specific core from background tasks of the operating system (through the Linux command \emph{shield}), and set the CPU governor to \emph{performance}, such that the core is always at its highest frequency (in our case 4.2 GHz). 

\subsection{Execution Time}
\label{sec:exp_exec_time}

\begin{figure}[t!]
\centering
\includegraphics[width=0.45\textwidth]{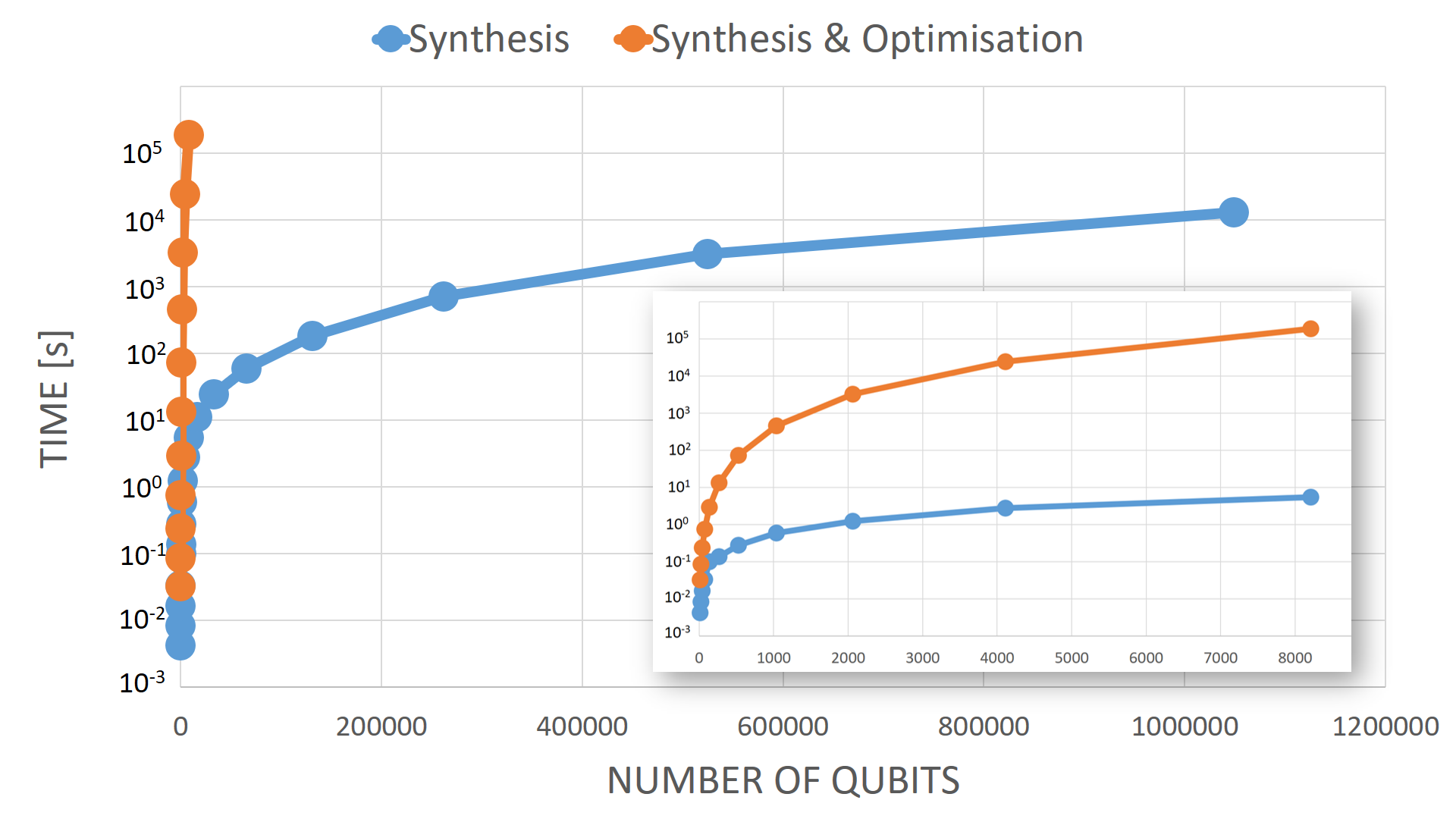}
\caption{Execution times for the "Synthesis" and the "Synthesis \& Optimisation" scenarios. The logarithmic Y-axis represents the execution time in seconds, and the X-axis represents the total number of qubits/wires in the resulting circuit. The encapsulated box provides more details on the first part of the global chart.}
\label{fig:res01}
\end{figure}

\begin{figure}[t!]
    \centering
    \includegraphics[width=0.45\textwidth]{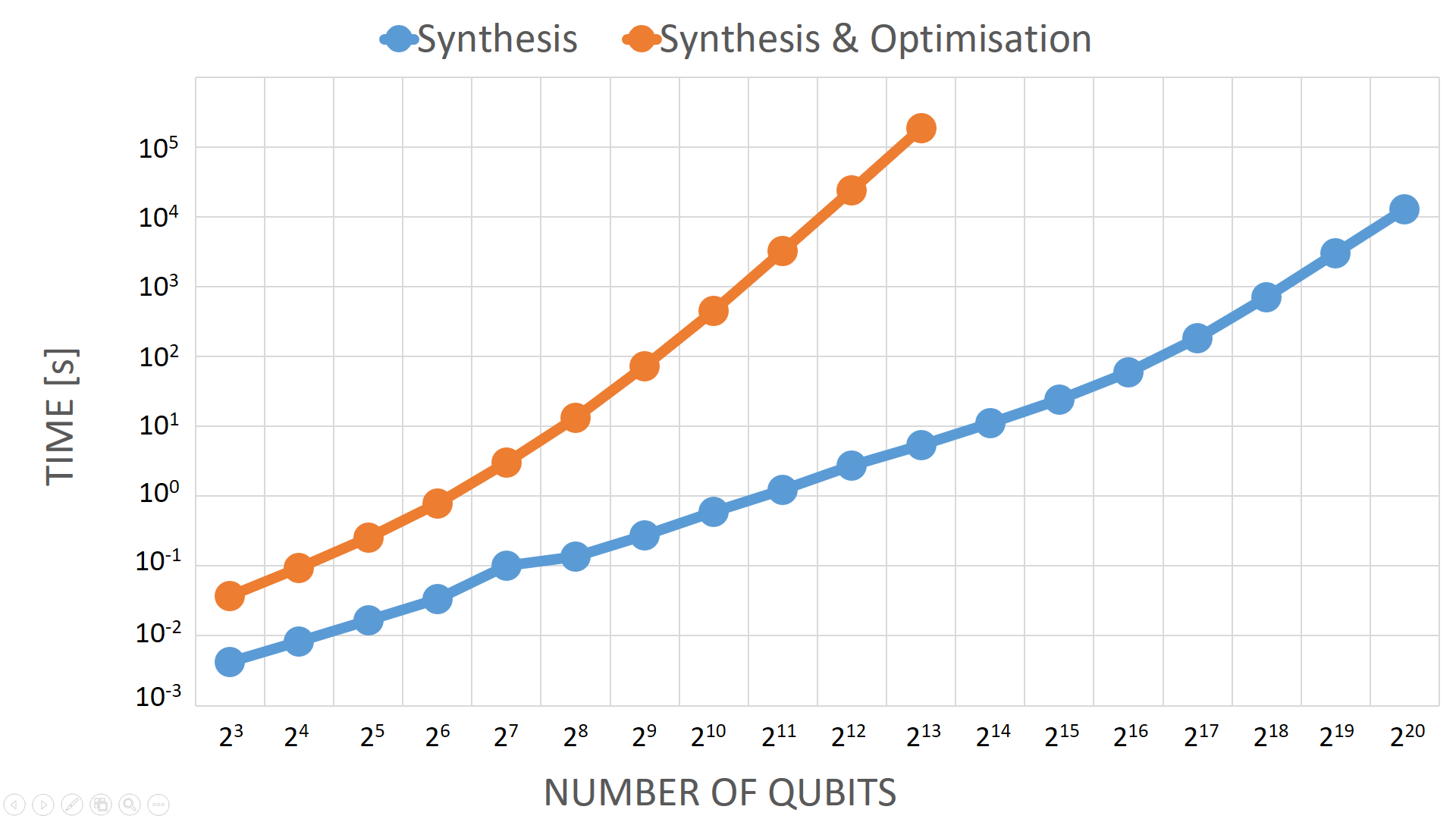}
    \caption{Execution times for the "Synthesis" and "Synthesis \& Optimisation" scenarios. The Y-axis and the X-axis represent the execution time in seconds and the total number of qubits/wires in the resulting circuits respectively.}
    \label{fig:res02}
\end{figure}

Figs.~\ref{fig:res01} and ~\ref{fig:res02} illustrate the results of the carried out experiments for the case of "Synthesis" (blue line) as well as "Synthesis \& Optimisation" (orange line). The X-axis represents the total number of qubits $q$ (e.g. input address qubits+ancillae) in the circuit, such that it is presented in linear and logarithmic scales for Figs.~\ref{fig:res01} and ~\ref{fig:res02} respectively. The Y-axis represents the execution time (in seconds), such that it is presented in logarithmic scale. The subfigure in Fig.~\ref{fig:res01} illustrates the results obtained when $2\leq n\leq 9$.   

As mentioned previously, "Synthesis" is faster than "Synthesis \& Optimisation", and as a result Figs.~\ref{fig:res01} and~\ref{fig:res02} include more sample points for the "Synthesis" scenario. We observed that for the "Synthesis" step (see Figs.~\ref{fig:res01} and~\ref{fig:res02}), QUANTIFY is satisfyingly responsive, taking less than a second for a total number of qubits $q$ between 15 and 1038 (e.g. $2\leq n\leq 9$). It had an execution times of 1.26, 2.52 and 5.13, 11.1, 24.24, 59.31, 181.61, 718.57, 3039.98, and 12782.44  seconds for the case of $n=10,...,19$ respectively. With respect to the combination of "Synthesis \& Optimisation" steps (see Figs.~\ref{fig:res01} and~\ref{fig:res02}), QUANTIFY executed in a very short amount of time, with less than a second for number of qubits $q$ between 15 and 74 (e.g. $2\leq n\leq 5$). The execution time increases linearly with respect to the number of qubits, with values of 2.92, 13.2, 72.66, 457.38, 3252.13, 24459.8 and 186536.9 seconds for the case of $n=6,...,12$ respectively.
 
There is strong evidence that the reason behind the high execution times is: a) the lack of specialised data structures in Cirq, and b) the missing parallelisation. We conjecture that straightforward optimisations will have significant speed-ups.

\subsection{Memory Usage}
\label{sec:exp_mem_usage}

\begin{figure}
    \centering
    \includegraphics[width=0.4\textwidth]{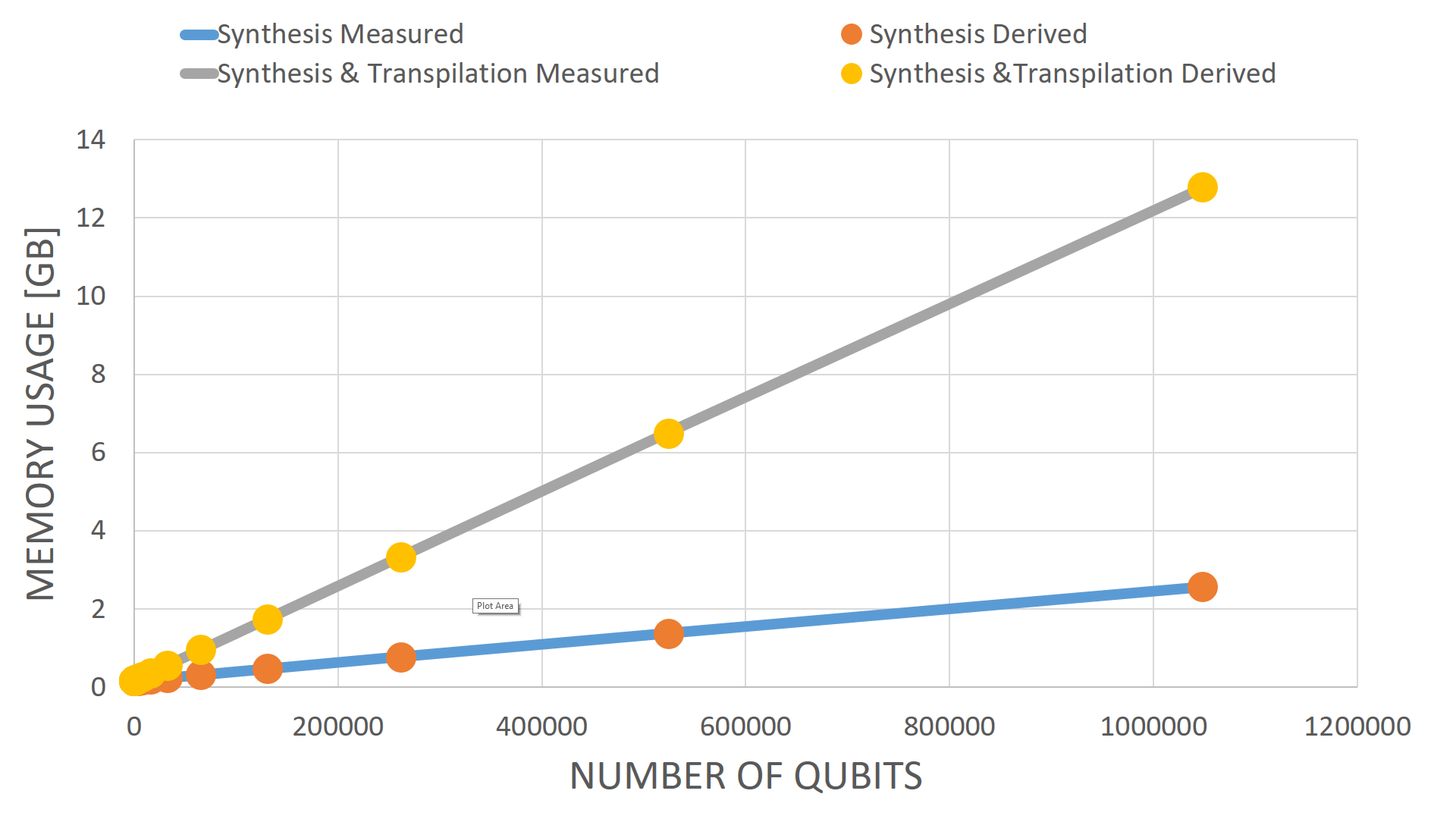}
    \caption{Comparison between the measured and derived models of the memory usage of QUANTIFY for large number of qubits by considering both "Synthesis" and "Synthesis \& Transpilation" scenarios.}
    \label{fig:memory_footprint}
\end{figure}

Fig.~\ref{fig:memory_footprint} illustrates the memory usage of QUANTIFY. The X-axis represents the total number of qubits $q$ in the circuit for $2\leq n\leq 19$. The Y-axis represents the memory usage (in GB). The experiments were carried out for the case of "Synthesis" and "Synthesis \& Transpilation" steps in order to assess the memory footprint of QUANTIFY. During Transpilation, the Toffoli gates of the benchmarked QRAM circuits were decomposed into Clifford+T.

\begin{figure}
    \centering
    \includegraphics[width=0.4\textwidth]{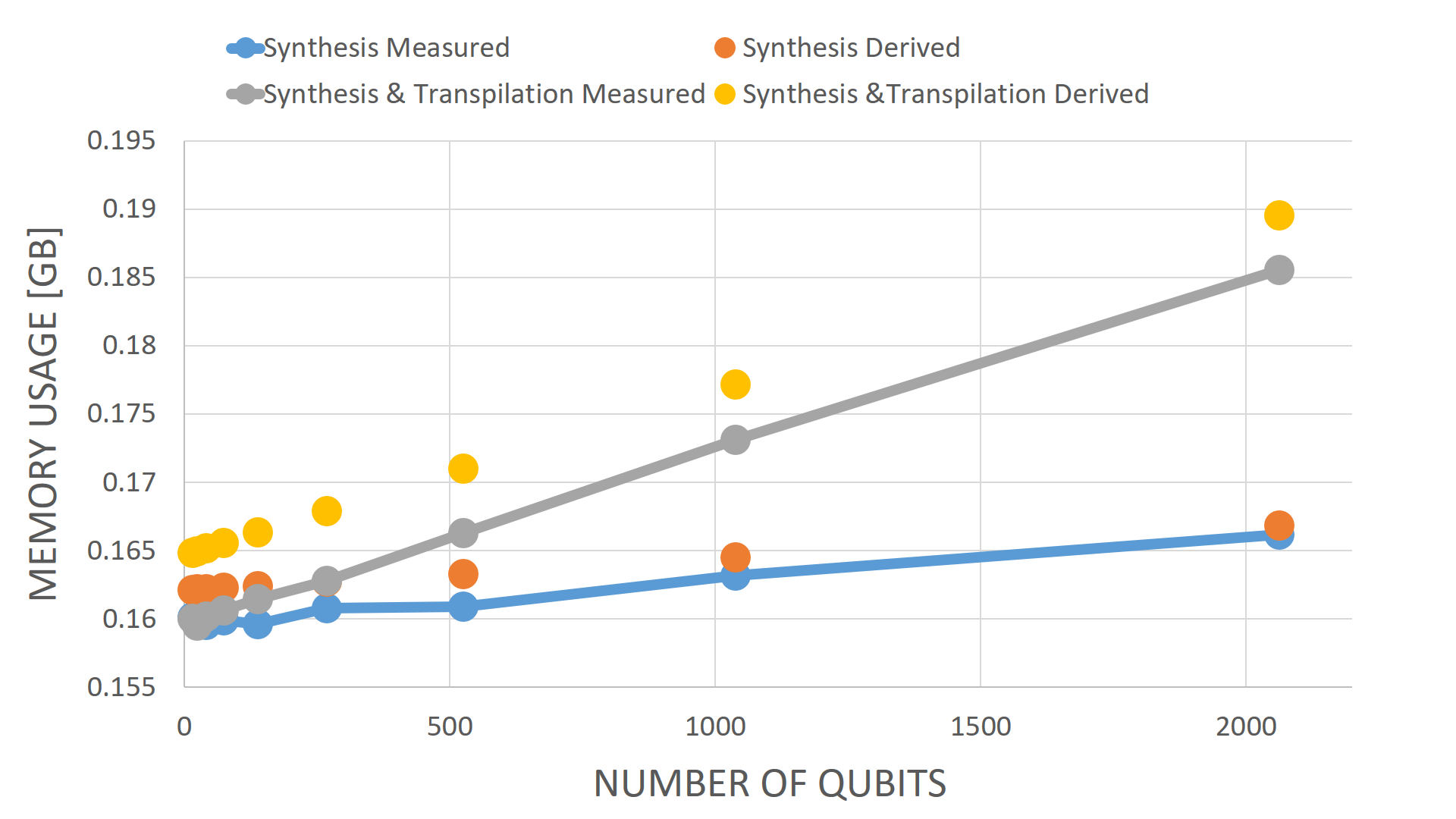}
    \caption{Comparison between the measured and derived models of the memory usage of QUANTIFY for small number of qubits by considering both "Synthesis" and "Synthesis \& Transpilation" scenarios.}
    \label{fig:memory_footprint_partial}
\end{figure}

From the obtained experimental results, for each of the case study we derived a model by fitting a linear function to the measured memory footprint. The accuracy of the derived models is illustrated in Figs.~\ref{fig:memory_footprint} and ~\ref{fig:memory_footprint_partial} both for the case of "Synthesis" (blue line (measured) vs red points (derived)) and "Synthesis \& Transpilation" (gray line (measured) vs yellow points (derived)) steps. For large number of qubits, our derived models are accurate. We ensure that the derived models reflect the reality accurately by calculating the root mean squared error: $4.866\times10^{-3}$ for the "Synthesis", and $1.431\times10^{-2}$ for "Synthesis \& Transpilation".

There is also a constant memory overhead of 0.16 GB. We speculate that this overhead is due to Cirq and its loaded libraries. Finally, we noticed that the amount of information generated for the "Synthesis \& Transpilation" step is about 5 times more than that of only "Synthesis" one. This factor may be due to the Toffoli transpilation to the Clifford+T.

\section{Conclusion}
\label{sec:conclusion}

QUANTIFY is a framework for the analysis and verification of quantum circuits. It is open sourced and based on Google Cirq. QUANTIFY includes flexible and novel optimisation methods, and it supports also the preparation and analysis of quantum circuits compatible with the surface code error correction. We benchmarked the performance of QUANTIFY using QRAM circuits. At the same time, for additional benchmarking purposes, the framework includes arithmetic circuits. QUANTIFY can within seconds synthesise and optimise circuits with thousands of qubits. Future work will include expanding and adding new optimisation and analysis capabilities, as well increasing the scalability and robustness of the framework. QUANTIFY will include all the necessary tools for realistic resource estimation of surface code protected quantum circuits.

\balance
\bibliographystyle{IEEEtran}
\bibliography{IEEEabrv,__main} 

\end{document}